\documentclass{article}
\usepackage{graphicx}      
\usepackage{varwidth}
\usepackage{setspace}
\usepackage{amsmath,bbm,amssymb,mathtools}
\usepackage{supertabular}
\usepackage{array}
\usepackage{url,float}
\usepackage{psfrag}
\usepackage{epstopdf}
\usepackage[width=.75\textwidth]{caption}
\DeclareCaptionType{copyrightbox}
\usepackage{ifthen}
\usepackage{lscape}

\newtheorem{keyword}{Keywords}
\usepackage[square,sort,comma,numbers]{natbib}

\usepackage{subcaption}
\usepackage{tikz}
\usepackage{pgfplots}
\usetikzlibrary{arrows,automata}
\pgfplotsset{compat=newest}
\usetikzlibrary{intersections}
\usetikzlibrary{plotmarks,automata}
\usepackage{balance}

\urldef{\mailsa}\path|{nathalie.cauchi, aabate}@cs.ox.ac.uk|

\newcommand{\po}{\mathbb{P}}     
\newcommand{\Rl}{\mathbb{R}}     

\newcommand{\eps}{\varepsilon}
\newcommand{\epsdel}{(\eps,\delta)}

\newcommand{\Cw}{C_{pw}}
\newcommand{\Cpa}{C_{pa}}

\newcommand{\Cahu}{C_{a}}

\newcommand{\Cni}{C_{w_{jn}}}
\newcommand{\Cnj}{C_{w_{jw}}}
\newcommand{\Cz}{C_{z_i}}

\newcommand{\Ben}{B_{en}}

\newcommand{\mahu}{m_{a}}

\newcommand{\Pout}{P_{rad_i}}

\newcommand{\Qr}{Q_{rw,r_i}}
\newcommand{\Qa}{Q_{rw,a_j}}
\newcommand{\Qo}{Q_{occ_i}}
\newcommand{\Qs}{Q_{solar_{jw}}}
\newcommand{\Qsa}{Q_{sa_i}}

\newcommand{\Ro}{R_{out}}

\newcommand{\Rni}{R_{ij}}

\newcommand{\Trwr}{T_{rw,r_i}}
\newcommand{\Trwa}{T_{rw,a}}
\newcommand{\dTrwr}{d{T}_{rw,r_i}}
\newcommand{\dTrwa}{d{T}_{rw,a}}

\newcommand{\Tz}{T_{z_i}}
\newcommand{\Tm}{T_{d}}
\newcommand{\dTz}{d{T}_{z_i}}

\newcommand{\Tni}{T_{w_{jn}}}

\newcommand{\dTni}{dT_{w_{jn}}}

\newcommand{\Tswb}{T_{sw,b}}
\newcommand{\dTswb}{d{T}_{sw,b}}

\newcommand{\Trwb}{T_{rw,b}}

\newcommand{\Tsaa}{T_{sa_i}}
\newcommand{\dTsaa}{d{T}_{sa_i}}

\newcommand{\To}{T_{out}}

\newcommand{\taub}{\tau_{sw}}

\newcommand{\rhow}{\rho_{h}}

\newcommand{\kb}{k_b}

\newcommand{\UAa}{(UA)_{a}}
\newcommand{\Uar}{(UA)_{r_i}}

\newcommand{\sigswb}{\sigma_{sw}}
\newcommand{\sigrwb}{\sigma_{rw,r_i}}

\newcommand{\sigz}{\sigma_{z_i}}

\newcommand{\siga}{\sigma_{sa_i}}
\newcommand{\sigrwa}{\sigma_{rw,a}}

\newcommand{\Va}{V_{a}}
\newcommand{\Vr}{V_{r_i}}

\newcommand{\wahub}{w_{a}}

\newcommand{\wrad}{w_{r_i}}

\newcommand{\Xbahu}{X_{a}}

\newcommand{\N }{\mathcal{N}}

\definecolor{mycolor1}{rgb}{0.00000,0.44700,0.74100}%
\definecolor{mycolor2}{rgb}{0.85000,0.32500,0.09800}%
\definecolor{mycolor3}{rgb}{0.92900,0.69400,0.12500}%
\definecolor{mycolor4}{rgb}{0.49400,0.18400,0.55600}%
\definecolor{mycolor5}{rgb}{0.46600,0.67400,0.18800}%
\definecolor{mycolor6}{rgb}{0.30100,0.74500,0.93300}%
\definecolor{mycolor7}{rgb}{0.63500,0.07800,0.18400}%

\begin{document}
	\title{Title of the paper}
	
	\footnotesize\date{}
	
	\author{Nathalie Cauchi, Alessandro Abate\\
		\footnotesize 	Department of Computer Science, 
		University of Oxford,
		Oxford, U.K\\
		\footnotesize \texttt{name.surname@cs.ox.ac.uk} \\ }
	
	\title{{Benchmarks for cyber-physical systems: \\ 
			A modular model library\\ for building automation systems}\\ (Extended version)}

\maketitle


\begin{abstract}                
Building Automation Systems (BAS) are exemplars of Cyber-Physical Systems (CPS), incorporating digital control architectures over underlying continuous physical processes. 
We provide a modular model library for BAS drawn from expertise developed on a real BAS setup. 
The library allows to build models comprising of
either physical quantities or digital control modules.
The structure, operation, and dynamics of the model can be complex, incorporating 
(i) stochasticity,  
(ii) non-linearities, 
(iii) numerous continuous variables or discrete states,  
(iv) various input and output signals, 
and (v) a large number of possible discrete configurations. 
The modular composition of BAS components can generate useful CPS benchmarks.    
We display this use by means of three realistic case studies, where corresponding models are built and engaged with different analysis goals. The benchmarks, the model library  and data collected from the BAS setup at the University of Oxford, are kept on-line at \url{https://github.com/natchi92/BASBenchmarks}
\end{abstract}

\begin{keyword}
cyber-physical systems, building automation systems, thermal modelling, hybrid models, simulation, reachability analysis, probabilistic safety, control synthesis 
\end{keyword}

\section{Introduction}

This paper describes a library of models for Building Automation Systems (BAS), 
which can be employed to create benchmarks for verification, control synthesis, or simulation purposes of Cyber-Physical Systems (CPS). 
The models are inspired by and built around an experimental setup within the Department of Computer Science at the University of Oxford, 
which is part of on-going research in collaboration with service engineers and industrial partners in the sector. 
This library allows to create numerous meaningful models for BAS, 
which are examples of CPS integrating continuous dynamics and discrete modes. 

Interest in BAS, also colloquially known as {\it smart buildings}, is gaining rapid momentum, 
in particular as a means for ensuring thermal comfort~(\cite{belkhouane2017thermal}), minimising energy consumption~(\cite{yu2015application,Soudjani2015:IEEE}), 
and ascertaining reliability~(\cite{Zhang2017178,FMTCauchi17}). 
Quantitative models are needed to evaluate system performance, 
to verify correct behaviour, 
and to develop specific control algorithms. 
An overview of the different BAS modelling techniques used in literature is presented in~\cite{Privara2013:EB}. 
Several simulation tools (see~\cite{crawley2008contrasting}) have been devised to aide in the development and analysis of models for BAS.
Attempting a multi-dimensional characterisation of the broad spectrum of existing BAS models,  
we can find either deterministic or stochastic ones, 
low- to high-dimensional ones,  
with discrete or continuous inputs and states.
The choice of a model is an art and a craft~(\cite{Ma2012:IEEE_Mag}):   
one must select simplifying assumptions that accurately reflect the operational performance of the BAS in specific real-world environments, 
and introduce uncertainty to represent un-modelled components, unknown parameters or random occupants. { We therefore aim to 
	simplify the modelling process such that simulation, verification or strategy synthesis can be carried out seamlessly.  
	Different verification and policy synthesis tools exist in literature~(\cite{dragomir2017refinement,hahn2013compositional,holub2016efficient}). They are typically specific to a particular type of model structure and in the case of stochastic or hybrid models are  oftenlimited to a small number of continuous variables. The use of such tools also requires expert knowledge on the specific formalism the tool makes use of.  
}

In order to display the versatility of the library of BAS models, 
we present three case studies that are built from its components. 
We focus on modelling temperature dynamics, 
a key element for ensuring thermal comfort.  
We employ the three generated models for different analysis goals, comprising simulation, reachability, and control synthesis. 
{The models and delineation of the case-studies are kept on-line at
	
	\begin{center}
	\url{https://github.com/natchi92/BASBenchmarks}.
	\end{center}
	
 This is to allow their use or modification for different applications and for comparison with other modelling approaches in BAS.  
	The repository also contains real data gathered from the BAS lab at Oxford, which can be employed for further modelling studies.
}
This article has the following structure: 
Section~\ref{sec:BAS} introduces the BAS modelling framework for CPS. 
We identify three modelling trade-offs that introduce different complexities on the model dynamics. 
Based on these trade-offs, we develop and analyse three case studies 
in Section~\ref{sec:CaseStud}.


\section{Building Automation Systems}
\label{sec:BAS}

\subsection{BAS: structure and components} 
\begin{figure*}[ht!]
	\centering
		\caption{Building automation system setup} 
	\begin{subfigure}{0.87\columnwidth}
		\centering
		\includegraphics[width=0.8\columnwidth]{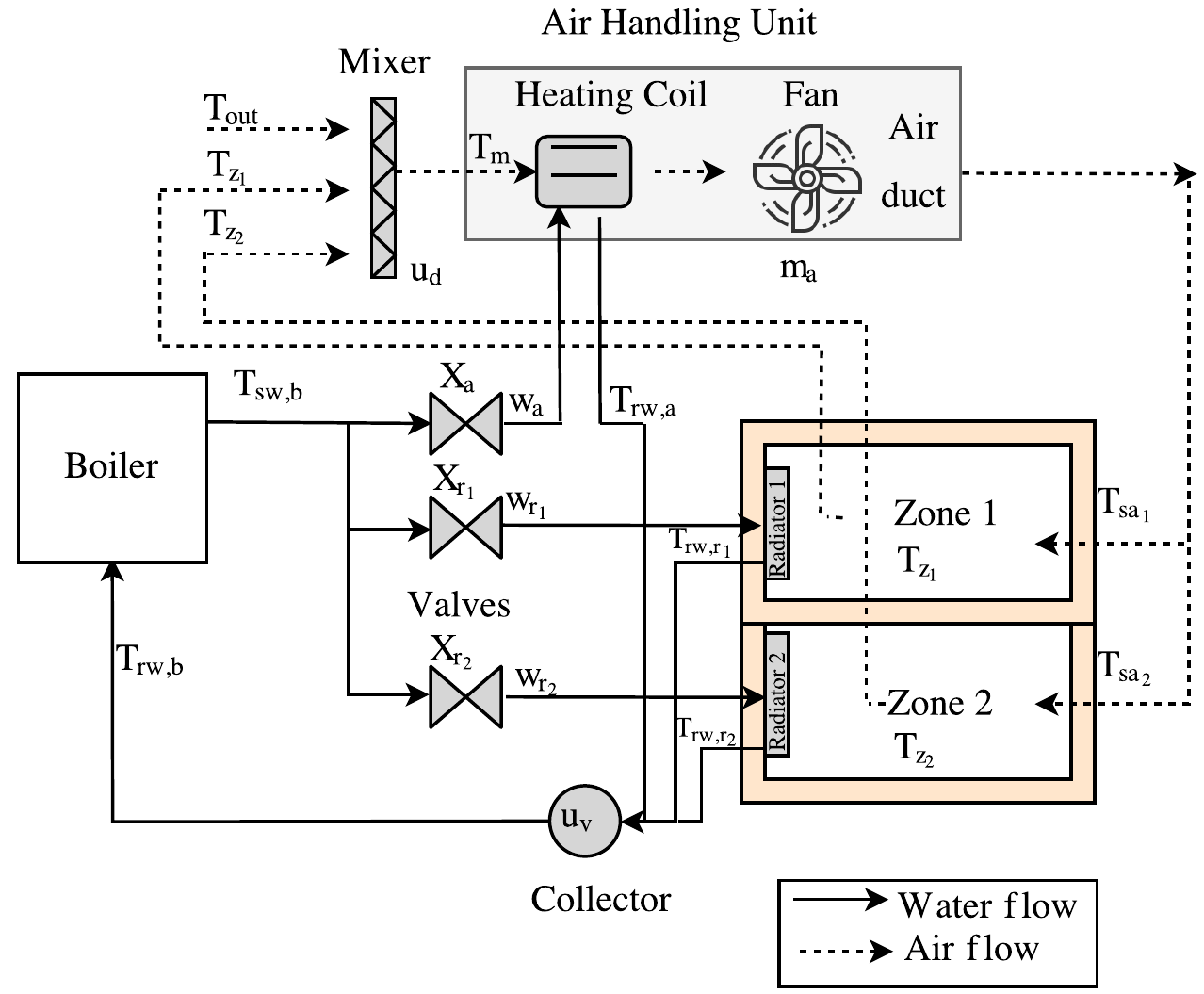}
		\caption{2-zone boiler-based heating system with air handling unit and radiators}\label{fig:Examples:1}
		\label{fig:cs1}
	\end{subfigure}\vspace{0.5cm}
	\begin{subfigure}{0.87\columnwidth}
				\centering
		\includegraphics[width=0.8\columnwidth]{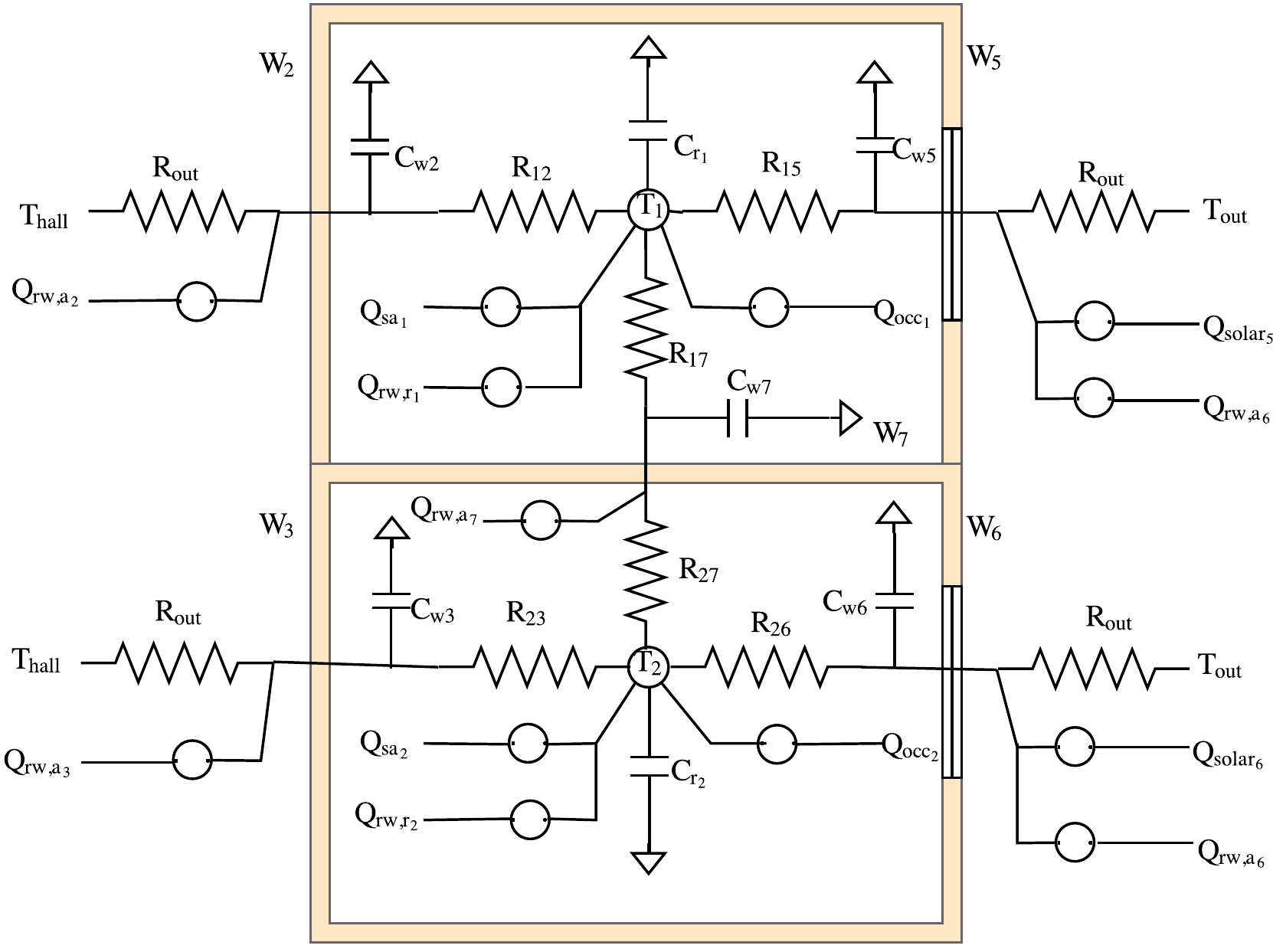}
		\caption{Resistance-capacitance circuit for the internal thermal dynamics within the two zones} 
		\label{fig:Examples:2}
	\end{subfigure}

\end{figure*}
BAS models clearly depend on the size and topology of the building~(\cite{Woohyun2017}), 
and on its climate control setup. 
In this work, 
we consider the BAS setup in the Department of Computer Science, 
at the University of Oxford.
A graphical depiction is shown in Figure~\ref{fig:cs1}. 
The BAS consists of two teaching rooms that are connected to a boiler-heated system. 
The boiler supplies heat  
to the heating coil within the AHU and to two radiators. Valves control the rate of 
water flow in the heating coil 
and in radiators. 
The AHU supplies air
to the two zones, which 
are connected back to back, 
and 
adjacent both to the outside and to an interior hall (cf. Figure~\ref{fig:Examples:1}). 
The zone air of both rooms can mix with the outside air
and exchanges circulating air with the AHU. 
Return water from the AHU heating coils and radiators is collected and pumped 
back to the boiler. Figure~\ref{fig:Examples:2} presents the Resistor Capacitance (RC) network circuit of the two zones~(\cite{peva}), 
which underpins the dynamics for temperature in the zone component - corresponding equations are in Table~\ref{eqn:models}. 
The heat level in each room is modified by 
(i) radiative solar energy absorbed through the walls, 
(ii) occupants, 
(iii) AHU input supply air, 
(iv) radiators, 
and (v) AHU return water. 
The effect of heat stored in the walls and in rooms is depicted with capacitors, 
whereas thermal resistance to heat transfer by the walls is depicted by resistor elements. 

\newpage

\subsection{BAS: dynamics and configurations} \label{subsec:BAS:models}

We define models for the individual components in the BAS system 
based on the expertise developed on the BAS setup at Oxford. 
Single components are intended as separate physical structures within the BAS. 
Their models are  
built from the underlying physics and are improved via industrial feedback and from existing literature~(\cite{ ,peva}). 
We obtain models with a number of unknown parameters: 
these are estimated and validated using data collected from the BAS setup~\cite{Kristensen2004:PES}.   
We list indices in Table~\ref{tab:index}, 
while all the quantities (variables, parameters, inputs) are listed in Table~\ref{tab:params}. 
Table~\ref{eqn:models} presents all the relations among variables in the model: 
algebraic relations define static couplings, 
whereas differential relations define the dynamics for the corresponding variables. 
The structure in Figure~\ref{fig:Examples:1}, 
the quantities in Table~\ref{tab:params}, 
and the variables (with associated dynamics) in Table~\ref{eqn:models}, 
together allow to construct global models for the complete BAS setup.  
We refer to the set of models describing the individual components (cf. Table~\ref{eqn:models}) as a library of models: 
one can select the individual components and models from the library, 
and build different BAS configurations. 
\begin{table}[h!]
	\centering	\caption{Indices}
	\resizebox{0.9\columnwidth}{!}{
		\begin{tabular}{|l|l||l|l|}
			\hline \textbf{Index} & \textbf{Reference} &\textbf{Index} & \textbf{Reference} \\ \hline \hline
			$a$ & AHU &$adj$ & adjacent zone\\ \hline
			$adj,out$& adjacent exterior zones & $b$ & boiler\\ \hline 
			$d$ & mixer & $hall$ & hallway \\ \hline 
			$i=\{1,2\}$ & individual zones & $jn=\{2,3,7\}$ &zone walls with no windows \\ \hline
			$jw=\{5,6\}$ & zone walls with windows & $j = jn \cup jw$&all zone walls \\ \hline 
			$l\in \{1,2\}$ & adjacent interior zone&	$occ$ & occupants \\ \hline
			$out$ & outside &	$r$ & radiator \\ \hline
			$ref$ & reference& $rw $ & return water \\ \hline 
			$sa$ & supply air&  $solar$ & solar energy \\ \hline 
			$sw$& supply water&$v$ & collector\\ \hline
			$w$ & wall &$z$ & zone\\ \hline 
			$h$ & water & $ar$& air \\ \hline
	\end{tabular}}
	
	\label{tab:index}
\end{table} 	
\begin{table*}[h!]
	\centering
		\caption{List of variables, inputs, and parameters}
		\begin{tabular}{|l|l|l|} \hline
			\textbf{Symbol} & \textbf{Quantity} & \textbf{Type}\\ \hline \hline
			$A_i$ & area of windows of each zone &constant\\ \hline
			$\Ben$& boiler switched off & discrete \\ \hline   
			$C$         &  capacitance           & constant \\ \hline
			$C_{pa}, C_{pw}$ & specific heat capacity of air and water & constant \\ \hline 
			$CO_{2_i}$& carbon-dioxide measurements in each zone & input\\ \hline
			$\kb$       & steady-state of the boiler       & constant \\ \hline 
			$m$         & mass air flow rate               & input \\ \hline $n$         & number of zones                  & constant \\ \hline
			$P_{out}$   & radiator rated power output      & constant$\backslash$input \\ \hline 
			$Q$ & heat gain & input \\ \hline 
			$R$         &  thermal resistance to heat & constant \\ \hline
			$T$         & temperature  & state$\backslash$input \\ \hline
			$u$         & mixing ratio                     & input 
			\\ \hline$(UA)$ & overall transmittance factor  & constant \\ \hline 	
			$V$         & volume                 & constant \\ \hline $w$ & water flow rate &  input\\ \hline 
			$w_{max}$   & maximum water-flow  & constant \\ \hline$X$ & valve position & input \\ \hline
			$\{\alpha, \beta,\mu\}$ & de-rating and offset factors & constants \\ \hline $\sigma$ & process noise & constant \\ \hline 			 
			$\rho$ & density  & constant \\ \hline $\tau$ &  time constant & constant \\ \hline 	
	\end{tabular}

	\label{tab:params}
\end{table*}

\begin{table*}[h!]
	\centering
		\caption{Components dynamics and functional relations among variables}
	\resizebox{\columnwidth}{!}{
		\begin{tabular}{|l|l|} \hline
			\multicolumn{2}{|c|}{\textbf{Component: Boiler}}\\ \hline	\hline		\textbf{Continuous variables} & \textbf{Relation} \\ \hline 
			$\begin{aligned}[t] 
			\dTswb(t)&= \begin{cases}
			0 & \Ben(t) = 0\\
			\left(\taub\right)^{-1}\left[ (-\Tswb(t) + \kb)dt\right] + \sigswb dW  & \Ben(t) =1
			\end{cases}
			\end{aligned}$	
			& differential \\ \hline 
			\multicolumn{2}{|c|}{\textbf{Component: Valve}} \\ \hline \hline
				\textbf{Continuous variables} & \textbf{Relation} \\ \hline 
			$\begin{aligned}[t]
			&	w(t)=\left(\tau\right)^{-1}\left[\exp(\ln(\tau)X(t))w_{{max}}\right]
			\end{aligned}$& algebraic \\[1.5ex] \hline 
			\multicolumn{2}{|c|}{\textbf{Component: Mixer}} \\ \hline \hline
				\textbf{Continuous variables} & \textbf{Relation} \\ \hline 
			$\begin{aligned}[t]
			&\Tm(t)=u_d\To(t) + (1-u_{d})(\sum_{i}\Tz(t))(n)^{-1}
			\end{aligned}$ & algebraic \\ \hline 
			\multicolumn{2}{|c|}{\textbf{Component: AHU heating coil}} \\ \hline \hline
				\textbf{Continuous variables} & \textbf{Relation} \\ \hline 
			$\begin{aligned}[t]
			&\dTrwa(t)= 
			\left({\Cw\rhow {\Va}}\right)^{-1} \left[({\Cw}\wahub(t)(\Tswb(t)-\Trwa(t))+{\UAa}(\Tm(t) -\Trwa(t) ))dt  \right] +\sigrwa dW
			\end{aligned}$& differential \\[1.5ex] \hline 
			\multicolumn{2}{|c|}{\textbf{Component: AHU air duct}} \\ \hline \hline
				\textbf{Continuous variables} & \textbf{Relation} \\ \hline 
			$\begin{aligned}[t]
			&\dTsaa(t)= 
			\left(\Cahu\rho_{a}{\Va}\right)^{-1}[\mahu(t)\Cpa(\Tm(t)-\Tsaa(t))+{\UAa}(\Tz(t) -\Tsaa(t) )]dt + \siga dW
			\end{aligned}$ & differential \\[1.5ex] \hline 
			\multicolumn{2}{|c|}{\textbf{Component: Radiator}} \\ \hline \hline	
				\textbf{Continuous variables} & \textbf{Relation} \\ \hline 		
			$\begin{aligned}[t]
			&\dTrwr(t)= 
			\left(\Cw\rhow {\Vr}\right)^{-1}\left[({\Cw}\wrad(t)(\Tswb(t)-\Trwr(t))+{\Uar}(\Tz(t) -\Trwr(t)))dt\right] + \sigrwb dW 
			\end{aligned}$ & differential \\[1.5ex] \hline 
			\multicolumn{2}{|c|}{\textbf{Component: Zone}} \\ \hline \hline
				\textbf{Continuous variables} & \textbf{Relation} \\ \hline 
			$\begin{aligned}[t]
			&\dTz(t)=
			\left(\Cz\right)^{-1}\left[\sum_{jn}\frac{\Tni(t)-\Tz(t)}{\Rni} + \Qr(t)  + \Qo(t) +\Qsa(t) \right]dt  + \sigz dW \\
			&\dTni(t) = \left(\Cni\right)^{-1}\left[\frac{{T_{adj,out}}(t) - \Tz(t)}{\Ro} + \sum_l \frac{{T_{adj_l}}(t)-\Tni(t)}{R_{lj}} + Q_{rw,a_{jn}}(t)   \right]dt + \sigma_{w_{jn}} dW\\
			&dT_{w_{jw}}(t) = \left(\Cnj\right)^{-1}\left[\frac{{T_{adj,out}}(t) - \Tz(t)}{\Ro} + \sum_l \frac{{T_{adj_l}}(t)-T_{w_{jw}}(t)}{R_{ljw}} + \Qs(t) + Q_{rw,a_{jw}}(t)   \right]dt + \sigma_{w_{jw}} dW\\
			&\Qr(t) = \Pout(\alpha_2(\Trwr(t) - \Tz(t)) + \alpha_1), \Qo(t)= \mu_i(CO_{2_i}(t)) + \beta_{1_i},\Qsa(t)= \mahu(t) \Cpa (\Tsaa(t) - \Tz(t)))  \\
			&\Qa(t) = \alpha_3(\Trwa(t) - T_{w_j}(t)), \Qs(t) = (\alpha_0A_i T_{out}(t) + \beta_2)\\
			\end{aligned}$
			& differential \\ \hline 
			\multicolumn{2}{|c|}{\textbf{Component: Collector}} \\ \hline \hline
				\textbf{Continuous variables} & \textbf{Relation} \\ \hline 
			$\begin{aligned}[t]
			&\Trwb(t)=u_{v}\Trwa(t)+ (1-u_v)(\sum_{i}\Trwr(t))(n)^{-1}
			\end{aligned}$ & algebraic \\[1.5ex] \hline 	
	\end{tabular}}

	\label{eqn:models} 
\end{table*}

A global model of the BAS set-up can be complex, 
comprising both algebraic and differential relations that may be  
further affected by process noise. A model also contains a number of 
inputs which can either be construed as control signals or as exogenous signals.
Some of the dynamics are non-linear in view of continuous variables that are bi-linearly coupled (cf. AHU air duct model in Table~\ref{eqn:models}).  
The number of continuous variables also increase substantially when considering a BAS setup with multiple zones: 
employing the zone component for a configuration with $n$ zones would result in a model with $(2n+1) + n$ continuous state variables. 
Furthermore, the model features multiple components that present switching discrete behaviours, 
affecting the dynamics of the continuous variables:  
these discrete modes are listed in Table~\ref{tab:modes} and  
can result in up to 144 discrete configurations.  
\begin{table}[h!] \centering
		\caption{Discrete operational modes} \label{tab:modes}
	\resizebox{.6\textwidth}{!}{
	\begin{tabular}{|c|c|}\hline
		\textbf{Component} & \textbf{Discrete Modes}\\ \hline \hline
		Boiler             & Boiler on, off ($\Ben$)\\ \hline
		AHU air duct       & Fan off,  medium, high ($\mahu$)\\ \hline
		Mixer              & Open, closed ($u_d$)\\ \hline
		AHU heating coil   & Valve healthy, faulty ($\Xbahu$)\\ \hline
		Radiator heating coil 1 & Valve healthy, faulty ($X_{r_1}$)\\ \hline
		Radiator heating coil 1  & Valve fully open, half open, closed ($X_{r_1}$)\\ \hline
	\end{tabular}}

\end{table}

In order to tackle the complexity of global BAS models and to add a level of flexibility to the modelling framework, 
we consider each BAS component as a separate module, characterised by inputs and output elements, 
and internal variables. 
We make use of individual modules describing component type, 
and then connect different modules based on possible physical couplings. 
Coupling between different modules is also achieved via input-output relationships: 
e.g., in the zone module we have coupling between two zones through the continuous variable $T_{adj,l}$ corresponding to the adjacent zones, 
which for the wall separating the two zones (cf. $W_7$ in Figure~\ref{fig:Examples:2}) corresponds to the individual zone temperatures of the two zone modules (cf. Table~\ref{eqn:models} zone equations).  
Having such a modular structure for the individual components provides an added level of versatility, 
since we can connect different components to create various new models. 
Modularisation 
also allows (i) to perform analysis of the whole setup by executing analysis of individual modules and (ii) to extend the library of models by defining new modules that connect to existing modules via their input-output relations. 
\subsection{BAS: description of model library}

The library of BAS components comes in the form of MATLAB scripts. Each script represents an individual BAS component. The models are in state-space form and are of two types linear or bilinear depending on the component they represent. They are defined using the symbolic toolbox, are parametrised and can be described both in discrete and continuous time. We  provide the parameters which we estimated from data gathered from the BAS lab to construct the individual models. However, users can easily make use of their own parameters and construct their own model. Different components can be connected together based on their input-output relations by cascading the different symbolic models for each component. Once this is done, the provided scripts allow you to simulate the models and plots for the defined output variables are presented. 
	
\section{Case studies}
\label{sec:CaseStud}
	Next we set up three case studies and present the trade off between the discussed elements of complexity. 
	For each of the case studies, 
	we (i) establish the dynamics of the models, (ii) how they are constructed from the library of components and (iii)  
	and describe the results obtained
\subsection{Two-zone heating setup with deterministic or stochastic dynamics}
\label{subsec:cs:1}

	\begin{figure}[h!]
		\psfragscanon
		\centering 
				\caption{BAS setup for the first case study}
		\includegraphics[width=0.8\columnwidth]{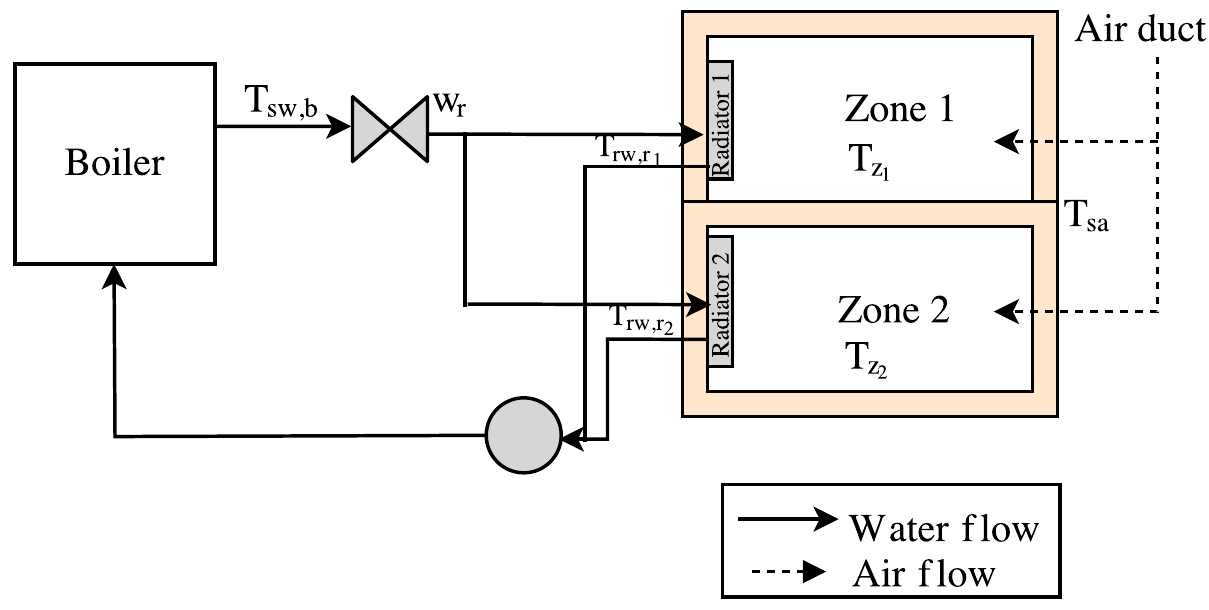}

		\label{fig:cs1:setup}
	\end{figure}

\begin{figure}[h!]
		\centering
		\psfragscanon
				\caption{First case study: simulations over two days}
		\input{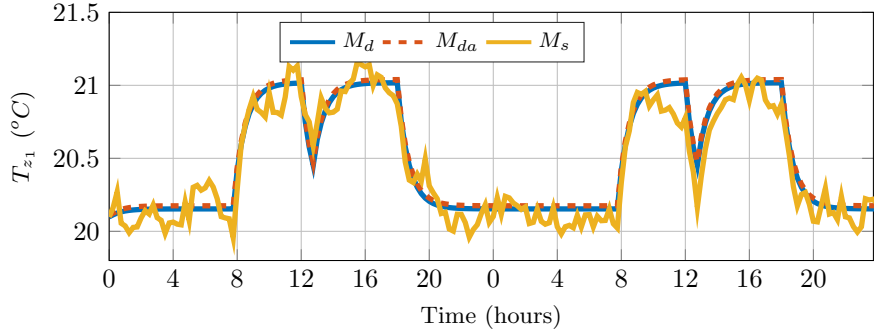}

		\label{fig:Cs1_sim}
	\end{figure}

    We consider two zones, each heated by one radiator and with a common supply air, 
    as portrayed in Figure~\ref{fig:cs1:setup}. 
	From  Table~\ref{eqn:models}, 
	we select two components and corresponding models: the radiator and the zone. 
	We simplify these models with the following assumptions: 
	(i) the  wall temperature is constant across the zones and is 
	a fixed value ($T_{w,ss}$); 
	(ii) the boiler is switched ON providing a supply temperature $T_{sw,b_{ss}}$; 
	(iii) we fix both the mass air flow rate $m_a$ and the radiator water flow rate $w_r$; 
	and  (iv) we do not include 
	the heat gain from the windows and 
	the AHU heating coils ($T_{w_{{ss}}}$) in each zone. 	
	We 
	obtain a model with the four state variables 
	$x^T = \begin{bmatrix}T_{z_1}& T_{z_2}&  T_{rw,r_1} & T_{rw,r_2} \end{bmatrix}^T$
	and with a common supply temperature $u =T_{sa}$ as an input.
	For this setup we further consider three different dynamics: 
	(i) a purely deterministic one; 
	(ii) a deterministic model with additive disturbance; 
	and (iii) a stochastic model. 
	For (i) and (ii), we thus remove the process noise in the template model, 
	while for models (i) and (iii) we do not include the occupancy heat gain. 
	We also discretise the dynamics by a Forward-Euler scheme (for the deterministic models) and a Euler-Maruyama scheme (for the stochastic model), 
	using a uniform sampling time $\Delta=$ 15 minutes,  
	and obtain a set of 
	linear discrete-time models.
	One should note that the models being considered are not fully observable since for all 
	the individual zone temperatures (variables of interest) are the only variables provided as outputs. 
	The dynamics of the deterministic model (i) are described by 
	\begin{equation}
	\label{mod:det}
		\textbf{M}_d :\begin{cases}
		x[k+1] &= A x[k] + Bu[k] + Q_d \\ 
		y_d[k]   &= \begin{bmatrix}
		1&0&0&0\\
		0&1&0&0\end{bmatrix} x[k], 
		\end{cases}
	\end{equation}
	where again the matrices are properly sized and constructed based on the models in Table~\ref{eqn:models} 
	and where,
	\begin{equation*}
		Q_{d} =\begin{bmatrix}\begin{split}
		\frac{T_{w_{{ss}} } \Delta}{C_{z_1}R_{1}}~~~& \frac{T_{w_{{ss}}} \Delta}{C_{z_2}R_{2}}&\frac{\Cw w_{r_1} \Delta}{\Cw \rhow V_{r_1,b}} T_{sw,b_{ss}}~~~& \frac{\Cw w_{r_2} \Delta}{\Cw \rhow V_{r_2,b}} T_{sw,b_{ss}}
		\end{split}
		\end{bmatrix}^T	
	\end{equation*}
	Here  $R_{i}$ is the mean resistance offered by the walls. 
	The deterministic model with additive disturbances is 
	\begin{equation}\label{mod:da}
		\textbf{M}_{da}:\begin{cases}
		x[k+1] &= \begin{split}
		A &x[k] + Bu[k] + F_{da}d_{da}[k]\\
		& + Q_{da} 
		\end{split}\\
		y_{da}[k]   &= \begin{bmatrix}
		1&0&0&0\\
		0&1&0&0\end{bmatrix}  x[k]. 
		\end{cases} 
	\end{equation}
	We have extended~\eqref{mod:det} with additive noise vector 
	$d_{da}^T = \begin{bmatrix}CO_{2_1}&CO_{2_2}\end{bmatrix}^T$
	(cf. $\Qo$ in Table~\ref{eqn:models}) representing the different $CO_2$ levels in each zone. 
	$Q_{da}$ is defined as
	$Q_{da} =  Q_d + \begin{bmatrix}\frac{\beta_{1_1}\Delta}{C_{z_1}}& \frac{\beta_{1_2}\Delta}{C_{z_2}}&0&0\end{bmatrix}^T$ 
	and $F_{da}$ is a properly sized matrix.
	The stochastic model is expressed by extending \eqref{mod:det} to include process noise, as 
	\begin{equation}\label{mod:s}
		\textbf{M}_s: \begin{cases}
		x[k+1] &=Ax[k] + Bu[k] + Q_{d} + \Sigma W[k] \\ 
		y_s[k]   &= \begin{bmatrix}
		1&0&0&0\\
		0&1&0&0\end{bmatrix} x[k], 
		\end{cases}
	\end{equation}
	where 
$
		\Sigma = diag([(\sqrt{\Delta}\sigma_{z_1})^2 \; (\sqrt{\Delta}\sigma_{z_2})^2 \;(\sqrt{\Delta}\sigma_{rw,r_1})^2 \;(\sqrt{\Delta}\sigma_{rw,r_2})^2])$
	encompasses the variances of the process noise for each state.  
	$W = \begin{bmatrix}w_1& w_2& w_3& w_4\end{bmatrix}^T$ are independent Gaussian random variables, 
	which are also independent of the initial condition of the process. 
	A simulation run of all three models is depicted in Figure~\ref{fig:Cs1_sim}.
	\subsubsection{Reachability analysis}
	\label{subsec:cs:1:analysis} 
	For this case study we would like to perform the following verification task: 
	to decide whether traces generated by the models remain within a specified safe set for 
	a given time period. 
	This is achieved by reachability analysis, 
	which takes a probabilistic flavour for the stochastic model. 
	The safe set is described as an interval around the temperature set-point $T_{SP} = 20^oC$ $\pm 0.5^oC$. 
	We constrain the input $u$ to lie within the set $\{T_{sa} \in \Rl  | 15 \le T_{sa} \le 22\}$ for all models and 
	employ a fixed time horizon $N = 6 \times \Delta = 1.5$ hours. 
	We perform reachability analysis with 
	Axelerator~(\cite{cattaruzza2015unbounded}),
	while we 
	use FAUST$^2$~(\cite{soudjani2015faust}) to perform probabilistic reachability analysis of $\mathbf{M}_s$. 
	
\begin{figure}[h!]
		\centering
		\psfragscanon
		\caption{First case study}
		\begin{subfigure}[b]{0.45\linewidth}
			\centering
			\resizebox{\columnwidth}{!}{
%
%
\definecolor{mycolor1}{rgb}{0.00000,0.44700,0.74100}%
\definecolor{mycolor2}{rgb}{0.85000,0.32500,0.09800}%
\definecolor{mycolor3}{rgb}{0.92900,0.69400,0.12500}%
\definecolor{mycolor4}{rgb}{0.49400,0.18400,0.55600}%
\definecolor{mycolor5}{rgb}{0.46600,0.67400,0.18800}%
\definecolor{mycolor6}{rgb}{0.30100,0.74500,0.93300}%
\definecolor{mycolor7}{rgb}{0.63500,0.07800,0.18400}%
\begin{tikzpicture}

\begin{axis}[%
width=2.5in,
height=1.3in,
at={(0,0)},
scale only axis,
xmin=10,
xmax=22,
ymin=5,
ymax=22,
axis background/.style={fill=white},
xmajorgrids,
ymajorgrids,
ylabel={\small $T_{z_2}$ ($^oC$)},
xlabel={\small$T_{z_1}$ ($^oC$)},
legend style={legend cell align=left, align=left, draw=white!15!black}
]
\addplot[fill=mycolor1, draw=black, forget plot] table[row sep=crcr] {%
	%
	11	6\\
	11	6\\
	12	6\\
	13	6\\
	14	6\\
	15	6\\
	15	6\\
	16	7\\
	16	7\\
	17	8\\
	17	8\\
	18	9\\
	18	9\\
	19	10\\
	19	10\\
	20	11\\
	20	11\\
	21	12\\
	21	12\\
	22	13\\
	22	13\\
	22	14\\
	22	15\\
	22	16\\
	22	17\\
	22	18\\
	22	19\\
	22	20\\
	22	21\\
	22	21\\
	21	21\\
	20	21\\
	19	21\\
	18	21\\
	17	21\\
	16	21\\
	15	21\\
	15	21\\
	14	20\\
	14	20\\
	13	19\\
	13	19\\
	12	18\\
	12	18\\
	11	17\\
	11	17\\
	11	16\\
	11	15\\
	11	14\\
	11	13\\
	11	12\\
	11	11\\
	11	10\\
	11	9\\
	11	8\\
	11	7\\
	11	6\\
};

\end{axis}

\end{tikzpicture}%
}
			\caption{Reach tube of $M_d$ over whole time horizon}\label{fig:MdRA}
		\end{subfigure}~
		\begin{subfigure}[b]{0.45\linewidth}
			\centering
				\resizebox{\columnwidth}{!}{
					\input{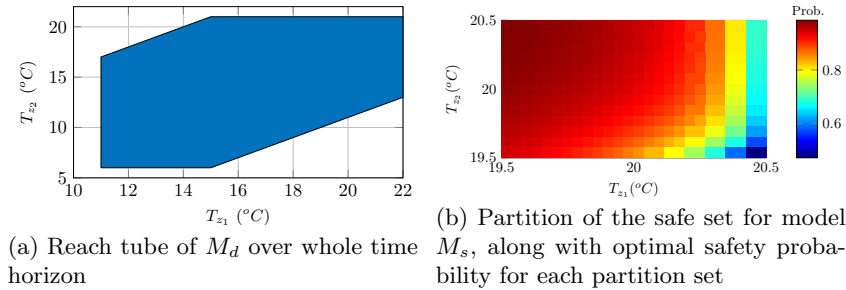}}
				\caption{Partition of the safe set for model $M_s$, 
					along with optimal safety probability for each partition set}\label{fig:PR}
		\end{subfigure}
	\end{figure}

	In order to perform reachability analysis using Axelerator, 
	for each of the models we set the initial condition as $\begin{bmatrix}T_{z_1}&T_{z_2}&T_{rw,r_1}&T_{rw,r_2}\end{bmatrix}^T$ = $\begin{bmatrix}
	18&18&35&35
	\end{bmatrix}^T$. 
	The reach tube for model $\mathbf{M}_{d}$ over the whole time horizon is shown in Figure~\ref{fig:MdRA}: it encompasses the union of reachable states over that horizon. 
	The results obtained using Axelerator (cf. Figure~\ref{fig:MdRA}) are conservative results and confirm that the model can indeed stay in the required safe set for some initial states,  but can also exit it.
	One can note the coupling between the two zones and that zone 1 tends to stay at higher zone temperatures then zone 2. 
	A similar reach tube is obtained for model $\mathbf{M}_{da}$.
	
	Similarly, we perform probabilistic reachability analysis on model $\mathbf{M}_s$ by defining the same safe set 
	and assuming an input set of $[15\;22]$.
	The resulting adaptive partition of the safe set along with the optimal safety probability for each partition set is depicted in Figure~\ref{fig:PR}. 
	When performing probabilistic reachability analysis using model $\mathbf{M}_s$ (cf. Figure~\ref{fig:PR}), we deduce that the models have a high probability of being within the required safe set, specifically to have $T_{z_1} \in [19.5\;20]$ and $T_{z_2} \in [19.5 \; 20.5]$.

	\subsection[Two-zone heating setup with large number of continuous variables]{\begin{varwidth}[t]{\linewidth}Two-zone heating setup with large number of \\continuous variables\end{varwidth}} 
	\label{subsec:cs:2}
	{In this second case study we focus on the dynamics of the zone component from 
		Table~\ref{eqn:models} and consider the two zones shown in Figure~\ref{fig:Examples:2}.
		We assume that (i) a central fan pumps in air in both rooms with a common supply temperature $15^oC \le T_{sa} \le 30^oC $, (ii) the input mass airflow $\mahu$ is fixed to $10\;m^3/hour$ and (iii) the return water temperature of the AHU heating coils is fixed ($T_{rw,a_{ss}} = 35^oC$).
		As in the previous case study, 
		the selected model is discretised using Forward-Euler, with a sampling time $\Delta = 15$ minutes, to obtain the discrete-time model 
		\begin{equation}
			\textbf{M}_{c}:
			\begin{cases}
			x_c[k+1] &= 
			A_{c}
			x_c[k] +B_cu_c[k]
			+ F_{c}d_c[k]+ Q_c
			\\
			y_c[k]   &= \begin{bmatrix}
			1&0&0&0&0&0&0 \end{bmatrix}x_c[k]. 
			\end{cases}
			\label{eqn:Mc}
		\end{equation}
		Here the variables are  
		\begin{equation*}
			x_c = \begin{bmatrix}T_{z_1} & T_{z_2}& T_{w_5}&T_{w_6}&T_{w_2}&T_{w_3}&T_{w_7}
			\end{bmatrix}^T,
		\end{equation*} 
		and a common fan supplies the two zones with a supply rate $u_c = T_{sa}$, 
		whereas 
		\begin{equation*}
		d_c = \begin{bmatrix}T_{out}& T_{hall}& CO_{2_1}&CO_{2_2}& T_{rw,r_1} &T_{rw,r_2}\end{bmatrix}^T,
		\end{equation*}
		and \begin{equation*}
			Q_c = \begin{bsmallmatrix} \frac{q_{c_0} \Delta}{C_{z_1}} &\frac{q_{c_0} \Delta}{C_{z_2}} & \frac{q_{c_2}\Delta}{C_{w_5}} & \frac{q_{c_2}\Delta}{C_{w_6}} & \frac{q_{c_1}\Delta}{C_{w_2}} &\frac{q_{c_1}\Delta}{C_{w_3}} 
			& \frac{q_{c_1}\Delta}{C_{w_7}} 
			\end{bsmallmatrix}^T,
		\end{equation*}
		where $q_{c_0} = \beta_{1_i} + \alpha_1 P_{rad_i}$, $q_{c_1} = \alpha_3 T_{rw,a_{ss}}$ and  $q_{c_2}= \alpha_0 A_2\beta_{2} +q_{c_1}$.
		Matrices $A_c,B_c,F_c$ are properly sized. 
		 Recall that the vector $d_c$ corresponds to the disturbance signals, while $Q_c$ represents constant additive terms within the model. We finally model the disturbances as random external effects.
		
		\subsubsection{Policy synthesis and refinement}
		\label{subsec:cs:2:analysis}
		
		For $\mathbf{M}_c$  we would like to synthesise a policy ensuring that the temperature within zone 1 does not deviate from the set point by more then 0.5$^oC$ over a time horizon equal to four hours (i.e $N= 16$). This requirement can be translated into the following PCTL property $\Phi:= \po_{= p} [\square^{\le N=16} |T_{z_1}-T_{SP}| \le 0.5]$. 		 
		We then aim at synthesising a policy maximising the safety probability $p$. This synthesis goal can be computationally hard due to the  number of continuous variables making up $\mathbf{M}_c$. To mitigate this limitation, we perform policy synthesis via abstractions~(\cite{HaesaertSA16}).
%
		We simplify~\eqref{eqn:Mc} into four abstract models using the technique in~(\cite{peva}). 
		The abstract models are labelled as $\mathbf{M}_{c_{a=\{4,\dots,1\}}},$ where $a$ represents the number of continuous variables of the corresponding abstract model.  
		The models take the form of Markov decision processes~(\cite{peva}). 
		We can quantify the error in the output variable, 
		which has been introduced by the different levels of abstractions, 
		through the use of \textit{$\epsdel$-approximate simulation relations}~(\cite{HaesaertSA16}). 
		The pair $\epsdel$ represents the deviation in the output trajectories between complex and abstract models and the differences in probability distribution of the processes, respectively.
		Such metrics allows the designer to select which of the considered abstract models provides the best trade off in precision:  
		it is desirable to achieve little deviation in both the output trajectories (small $\eps$) and in the probability distributions (small $\delta$). 

		\begin{table}[h!]
			\centering
			
			\caption[Second case study: error metrics $\epsdel$ for concrete and abstract models]
			{\begin{varwidth}[t]{\linewidth}Second case study: error metrics $\epsdel$ for  \\ concrete and abstract models.\end{varwidth}}
			\begin{tabular}{|c|ccccccc|}	
				\hline
				\textbf{$\delta$} 	& 1 & $10^{\frac{-1}{2}}$ & $10^{-1}$& $10^{\frac{-3}{2}}$& $10^{-2}$ & $10^{\frac{-5}{2}}$& $10^{-3}$\\  \hline \hline	
				$\mathbf{M}_{c_4}$ &0.0008&	0.1754&	0.2084&	0.2339&	0.2555&	0.2745&	0.2910\\ \hline			
				$\mathbf{M}_{c_3}$ &0.0006& 0.1933&	0.2312&	0.2598&	0.2831&0.3065&	0.3241\\ \hline	
				$\mathbf{M}_{c_2}$ &0.0011&0.1950&0.2373&0.2681&0.2928&	0.3155&	0.3278\\	\hline		
				$\mathbf{M}_{c_1}$ &0.0010&0.1953&	0.2371&0.2595&\underline{0.2854}&0.3103&	0.3254\\ \hline	
			\end{tabular}
		\label{tab:SimRel}
		\end{table}	
		
		We compute $\epsdel$-approximate simulation relations between $\mathbf{M}_c$ and the set of abstract models $\mathbf{M}_{c_{a=\{4,\dots,1\}}}$, 
		as presented in Table~\ref{tab:SimRel}. 
		The $\epsdel$ pair providing the optimal trade off 
		is obtained with the abstract model $\mathbf{M}_{c_1}$ and corresponds to ($0.2854,$ $ 10^{-2}$). 
		Next, we use FAUST$^2$ to perform a grid-based computation of the safety probability for $\mathbf{M}_{c_1}$ and obtain a model of size 14893 with an overall accuracy of $0.005$. 
		Over this approximation we synthesise the optimal policy for the abstract model which results in a safety probability of $p'= 0.9257$. 
		We refine the obtained policy~(\cite{peva}), which results in one that can be used with $\mathbf{M}_c$. 
		The overall process results in $\Phi$ being satisfied with a safety probability of $p = p'-\eta - N\delta = 0.7657$, where $\eta$ is the abstraction error introduced by FAUST$^2$. 
		The results obtained 
		further highlight
		that by 
		trading off the complexity in the number of continuous variables and computing $\epsdel$-simulation relations, we can synthesise policies using simpler models, 
		yet achieve high performance still when the refined policy is applied to the original model. 
		
		\subsection{Single zone heating with multiple switching controls}
		\label{subsec:cs:3}
		{In this third and last case study, we focus on mixer, AHU air duct, and zone components
		from Table~\ref{eqn:models}. 
		We select the AHU as the only	
		source of heat within the zone (the boiler is disconnected). 
		A pictorial description of this setup is in Figure~\ref{fig:cs3:bas}. 
		
				\begin{figure}[h!]
					\psfragscanon
					\centering
					\caption{Third case study: setup and model}
					\begin{subfigure}[b]{0.45\linewidth}
						\includegraphics[width=\columnwidth]{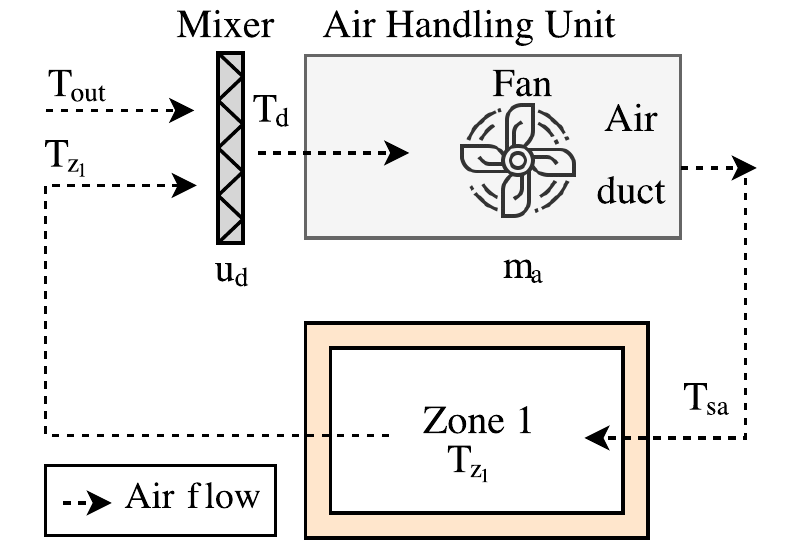}
						\caption{BAS setup}	
						\label{fig:cs3:bas}
					\end{subfigure}
					\begin{subfigure}[b]{0.45\linewidth}
						\psfragscanon
						\centering
						\resizebox{\columnwidth}{!}{
							\includegraphics[width=0.9\columnwidth]{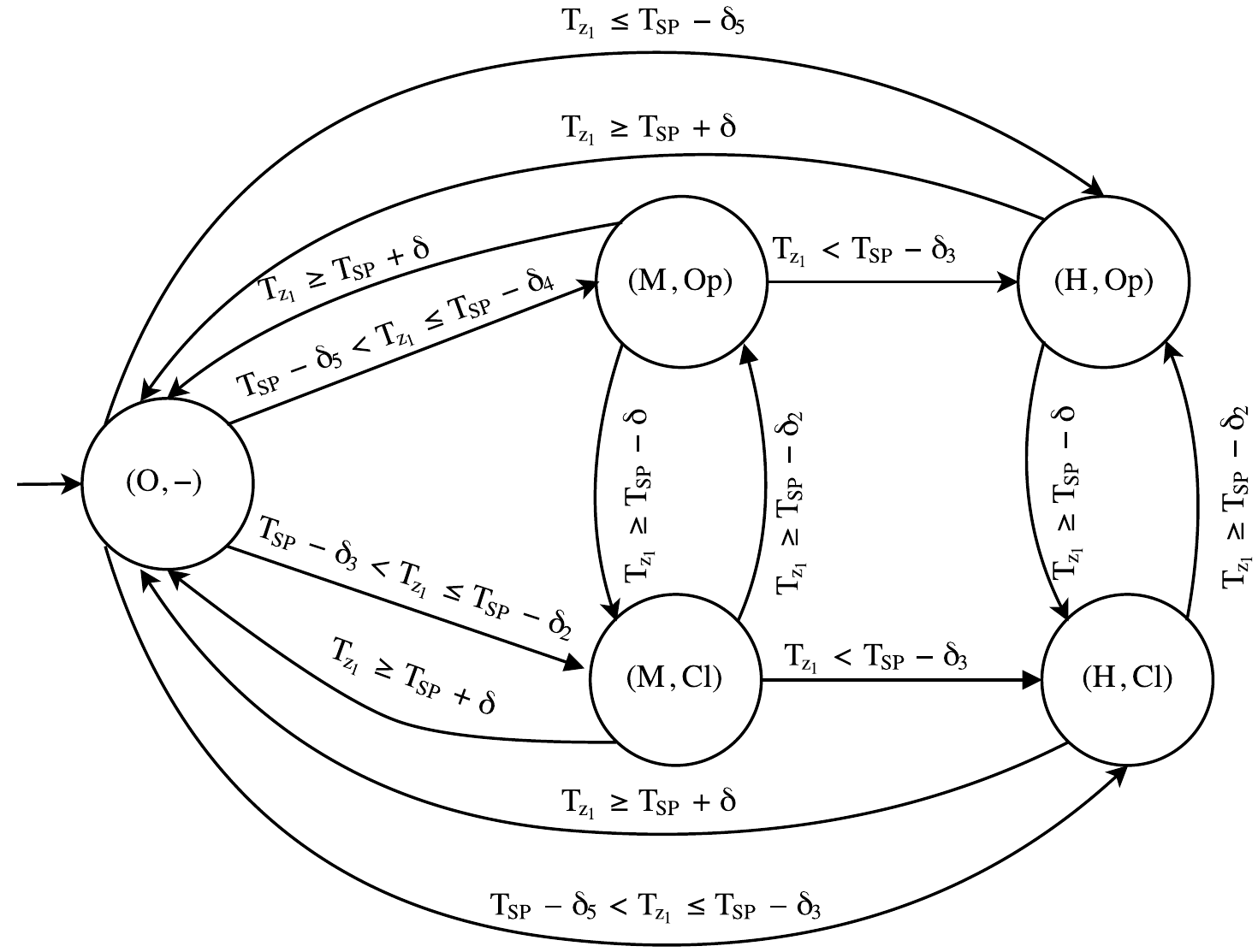}
						}	
						\caption{ Hybrid model showing discrete states and guard conditions; 
							the initial discrete state is $(O,-)$}
						\label{fig:mods:HS:a}
					\end{subfigure}
					
				\end{figure}
				
		The mixer operates in either of two modes: open ($Op$) or closed ($Cl$). 
		The AHU air duct recirculates air from either the internal zone (when $u_d = 0$ and the mixer is in mode $Op$) or from the outside (when $u_d =1$ and the mixer is in mode $Cl$) via the continuous variable $\Tm$ (output of the mixer component). The rate of air being pumped into the zone ($\mahu$) is controlled by the fan, which has three operating speeds 
		(off $O$, medium $M$, and high $H$). 
		The mixer position and the fan settings can be used to maintain a comfortable temperature within the zone. This setup can be described by a hybrid model. 
		The discrete modes $q$ are in the set $
		\{(O,-), (M,Op), (M,Cl), $ $(H,Op), (H,Cl)\},
			$
		and describe the possible configurations of fan operating speeds $\{O,M,H\}$ and mixer position $\{Op,Cl\}$. 
		(When the fan is switched off, the mixer can be in any position as no air is pumped into the zone.) 
		Continuous variables model the zone temperature $(T_{z_1})$ together with the supply air temperature $(T_{sa})$ being pumped into the zone. 
		Transitions between discrete modes are triggered by continuous dynamics crossing spatial guards: guards denote deviations from temperature set-point as $\delta$, $\delta_k,k=\{2,\dots,5\}$, $\delta <\delta_2<\delta_3<\delta_4<\delta_5$. 
		A graphical description of the overall hybrid model, 
		together with the different guard conditions,  
		is shown in Figure~\ref{fig:mods:HS:a}.  
		The continuous dynamics are built from Table~\ref{eqn:models} and follow  
			\begin{equation*}
				\begin{split}
				\dot T_{z_1} &= \left(C_{z_1}\right)^{-1} [\frac{T_{w_{ss}} -T_{z_1}(t)}{R}+ \mahu(t) \Cpa(T_{sa}(t) - T_{z_1}(t)+\mu_1 CO_{2_{1_{ss}}} + \beta_{1_1} ], \\
				\dot T_{sa} &= \left(\Cahu\rho_{a}{\Va}\right)^{-1} [ \mahu(t) \Cpa (\Tm(t) - T_{sa}(t)) + \UAa(T_{z_1}(t) - T_{sa}(t)].
				\end{split}
				\end{equation*}

			\noindent The variables $\mahu$ and $\Tm$ take values according to the discrete mode as 
			\begin{align*}
			\mahu(t) &= \begin{cases}
			0 & q(t) = (O,-), \\
			m_{a,med} & q(t)= (M,Op) \vee q(t) = (M,Cl),\\
			m_{a,high} & q(t) = (H,Op) \vee q(t) = (H,Cl),
			\end{cases} 
			\end{align*}
			and \begin{align*}
			\Tm(t) &=\begin{cases}
			T_{out} & q(t) = (M,Op) \vee q(t) = (H,Op),\\
			T_{z_1}(t) & \text{else.}\\
			\end{cases}
			\end{align*}
			Here, $m_{a,med}, m_{a,high}$ correspond to the air flow rates when the fan is operating at medium and high speeds. 
			
			\subsubsection{Reachability analysis}
			\label{subsec:cs:3:analysis}

			We are interested in performing reachability analysis of the hybrid model, 
			which we run 
			using SpaceEx~(\cite{frehse2011spaceex}). 
			Notice that in this case we do not discretise time and consider a continuous time horizon of $2$ hours. 
			We consider two different initial conditions:  
			in the first experiment we select an initial condition equal to $T_{z_1} = 15^oC$ and $T_{sa} = 15^oC$, 
			while in the second we set  $T_{z_1} = 20^oC$ and $T_{sa} = 20^oC$. 
			The resulting reach tube for the both experiments is shown in Figure~\ref{fig:cs3:reach:a}.
						
			\begin{figure}[ht!]
				\centering
								\caption{Third case study: reach tubes obtained from two different initial conditions} 
								\label{fig:cs3:reach}
				\begin{subfigure}{0.45\columnwidth}
					\includegraphics[width=0.9\columnwidth]{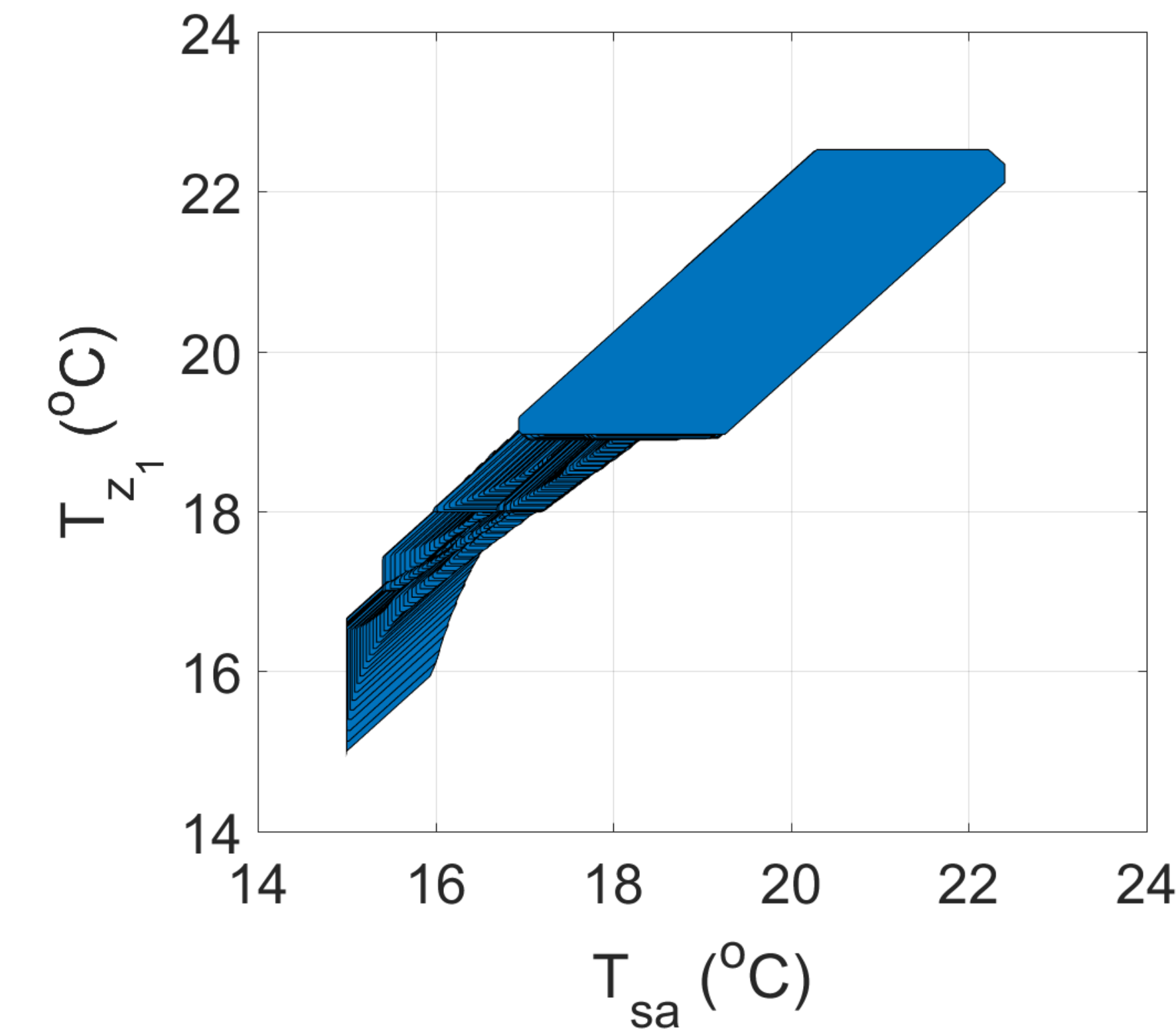}
					\caption{Initial condition: $T_{z_1} = 15^oC$, $T_{sa} = 15^oC$ }\label{fig:cs3:reach:a}
				\end{subfigure}~
				\begin{subfigure}{0.45\columnwidth}
					\includegraphics[width=0.9\columnwidth]{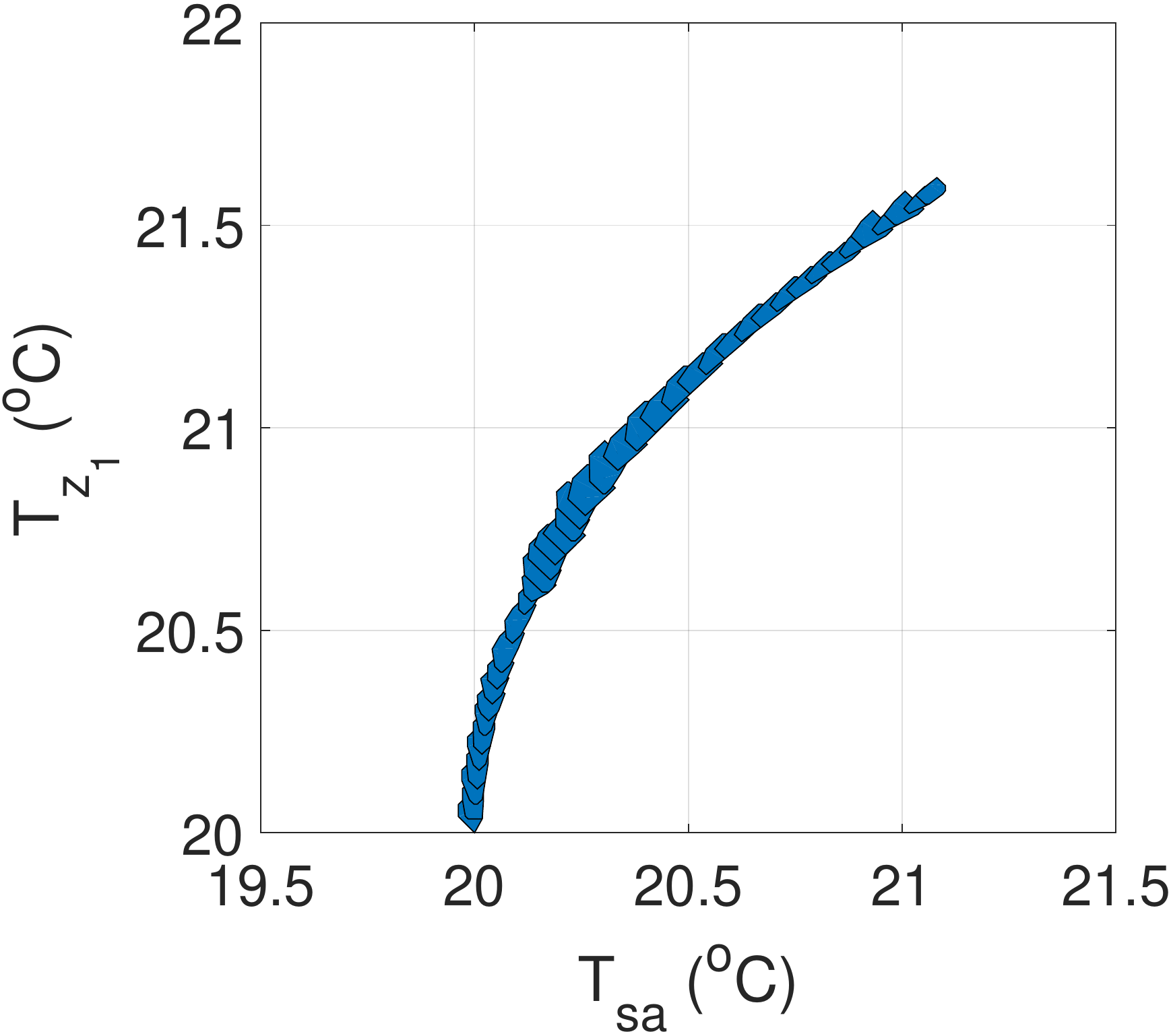}
					\caption{Initial condition:  $T_{z_1} = 20^oC$, $T_{sa} = 20^oC$ }
					\label{fig:cs3:reach:b}
				\end{subfigure}

			\end{figure}

			In Figure~\ref{fig:cs3:reach:a} we can see that the model initially is in state $(O,-)$ and jumps to a new state $(H,Op)$ such that warm outside air is pumped into the zone (due to the low temperature of the initial conditions). From  $(H,Op)$ it switches to $(M,Op)$ (notice the reduction in the gradient between the variables $T_{sa} \in [15 \; 18]$, $T_{z_1} \in [17\;19]$) and eventually switches back to $(O,-)$ in order to maintain the temperature within the comfort region. For Figure~\ref{fig:cs3:reach:b}, the reach tube shows that the system remains within the initial state $(O,-)$ over the whole time horizon. 
			
			\section{Conclusions}
			This paper has presented a library of CPS models for BAS. 
			The BAS modelling framework comprises three different types of complexities 
			(stochasticity, number of continuous variables, number of discrete modes) and is modular such that 
			various BAS components can be composed.  
			We illustrate the use of this BAS components library via three case studies, 
			each of which highlights a different side of the complexity trade off and solves a different problem 
			(simulation, (probabilistic) reachability analysis, and strategy synthesis respectively).  	
			Current work is 
			developed towards obtaining a compositional tool that is able to allow for easy construction of BAS models and for interfacing with different (i) verification tools for performing analysis and (ii) synthesis tools for computation of optimal strategies. 

			This work is to appear within the IFAC Conference on Analysis and Design of Hybrid Systems 2018 (ADHS 2018) \cite{Cauchi18ADHS}.

\section*{Acknowledgements}
This work has been funded by the {European Commission in the Seventh Framework Programme project AMBI} under Grant No.: {324432} and is in part supported by the Alan Turing Institute, UK and Malta's ENDEAVOUR Scholarships Scheme.
The authors would also like to thank Dario Cattaruzza, 
Sofie Haesaert and 
Honeywell Laboratories, Prague for their fruitful feedback.

\bibliography{Bib_rep}             
\bibliographystyle{plain}
\newpage

%
\section*{Appendix}
\subsection*{Discrete-time models for heating system of two zones with deterministic or stochastic dynamics}
\label{app:cs:1}
We present the corresponding system matrices of the three models $\mathbf{M}_{d}, \mathbf{M}_{da}, \mathbf{M}_s$ described in Sec.~\ref{subsec:cs:1}.
$\mathbf{M}_{d}$ is given by~\eqref{mod:det} and is characterised by the following system matrices:
\begin{align*}
&
A=\begin{bsmallmatrix}
0.6682&	0&	0.02632&	0\\
0&	0.6830&	0&	0.02096\\
1.0005&	0&	-0.000499&	0\\
0&0.8004&0&	0.1996\\
	\end{bsmallmatrix},~
	B=\begin{bsmallmatrix}
0.1320\\
0.1402\\
0\\
0\\
	\end{bsmallmatrix},~
	Q_{d}=\begin{bsmallmatrix}
3.4364\\
2.9272\\
13.0207\\
10.4166\\
	\end{bsmallmatrix}.
\end{align*} 
$\mathbf{M}_{da}$ is given by~\eqref{mod:da} and is characterised by the same system matrices used for $\mathbf{M}_d$ together with,
\begin{align*}
&F_{da}=\begin{bsmallmatrix}
8.760e-06&	0\\
0&	2.704e-07\\
0&	0\\
0&	0\\
\end{bsmallmatrix},~
Q_{da}=\begin{bsmallmatrix}
3.3378\\
2.9106\\
13.0207\\
10.4166\\
\end{bsmallmatrix}.
\end{align*}
$\mathbf{M}_{s}$ is given by~\eqref{mod:s} and is characterised by the same system matrices used for $\mathbf{M}_d$ together with,
\begin{align*}
&\Sigma=\begin{bsmallmatrix}
0.0774&	0&	0&	0\\
0&	0.0774&	0&	0\\
0&	0&	0.3872&	0\\
0&	0&	0&	0.3098\\
\end{bsmallmatrix}.
\end{align*}
In all cases we use,
\begin{align*}
T_{w_{ss}} = 18^oC,~&T_{SP} = 20^oC,~T_{sw,b_{ss}}= 75^oC,~
T_{z_{1_{ss}}}= 20^oC,\\
T_{z_{2_{ss}}}= 20^oC,~&T_{rw,r_{1_{ss}}}= 35^oC,~T_{rw,r_{2_{ss}}}=35^oC.\end{align*}
We perform a simulation run of all three models over a two-day period, 
with 
\begin{equation*}
	u[k] =\begin{cases}
	20^oC & \text{k is between 8am - 12pm, 1pm - 6pm},\\
	18^oC &  \text{otherwise},
	\end{cases}
\end{equation*}
and depict the resulting simulation runs for the temperature within zone 1 in Figure~\ref{fig:Cs1_sim}.

\subsubsection*{Reachability analysis using Axelarator}
\label{app:cs:1:axel}
To perform reachability analysis using Axelarator we need {to use the following commands in command line,
	\begin{align*}
	&\texttt{../aa/axelerator  Dynamics.txt } \texttt{-init}\\& \texttt{"[]" -sguard }\texttt{"[]" -inc -mpi -u} 
	\end{align*}
where the file \texttt{Dynamics.txt} contains the corresponding system matrices and defines the dimensions of both the state space and control inputs of the underlying model. The initial conditions are defined using \texttt{-init "[]"} and \texttt{-sguard "[]"} will force the tool to stay inside the safe region (if possible) defined within \texttt{"[]"}.  The commands \texttt{-inc -mpi -u} set the conditions of the problem such that Axelarator makes use of incremental search with multiple precision intervals and to process unsound results. 
We also provide Axelerator with the safe set defined in the form of a polyhedral, 
\begin{center}
$\begin{bmatrix}T_{z_1}\\T_{z_2}\\ \end{bmatrix}$  $>\begin{bmatrix} 19.5\\  19.5 \\ \end{bmatrix}$ $\wedge $ 
$\begin{bmatrix}T_{z_1}\\ T_{z_2}\\   \end{bmatrix}<$  $\begin{bmatrix}20.5\\  20.5 \\ \end{bmatrix}.$ 
\end{center}
This safe set is used as a reference once the resulting reach tube is computed.  
{The results obtained using Axelarator are in the form of polyhedral sets which define the reach tube of the underlying system over the N time steps. For $\mathbf{M}_d$ the polyhedral sets are given as,
	\begin{align*}
	\begin{bsmallmatrix}
		T_{z_1} & <& 22.2282, & T_{z_2}& <& 22,\\
		T_{rw,r_1}& < &40, & T_{rw,r_2} &< &40,\\
		-T_{z_1} &<& -10.0899,& -T_{z_2}&<& -5.40334,\\
		-T_{rw,r_1} &<& 105.958, & -T_{rw,r_2} &<& -11.1481, \\
		T_{z_1} + T_{z_2}& <& 44, & T_{z_1}+ T_{rw,r_1} &<& 62,\\
		T_{z_1}  + T_{rw,r_2} &<& 62, & T_{z_1} -T_{z_2} &<& 9.40198,\\
		T_{z_1}  - T_{rw,r_1}  &<& 116.097,& T_{z_1} -T_{rw,r_2} &<& 1.12788,\\
		-T_{z_1}  + T_{z_2}  &<& 7, &	-T_{z_1}  + T_{rw,r_1} &<& 25, \\
		-T_{z_1}  + T_{rw,r_2} &<& 25,&-T_{z_1} -T_{z_2} &<& -15.4932, \\
		-T_{z_1} -T_{rw,r_1} &<& 95.8203,&	-T_{z_1} -T_{rw,r_2} &<& -21.238,\\
		T_{z_2}  + T_{rw,r_1}& <& 62, & T_{z_2}  + T_{rw,r_2} &<& 62,\\
		T_{z_2}  -T_{rw,r_1} &<& 114.105, & T_{z_2}  -T_{rw,r_2} &<& -5.50237,\\
		-T_{z_2}  + T_{rw,r_1} &<& 25, &-T_{z_2}  + T_{rw,r_2} &<& 25,\\
		-T_{z_2}  -T_{rw,r_1} &<& 100.555,&-T_{z_2}  -T_{rw,r_2} &<& -16.5515,\\
		T_{rw,r_1}+ T_{rw,r_2} &<& 80, &	T_{rw,r_1} -T_{rw,r_2} &<& 10,\\
		-T_{rw,r_1}+ T_{rw,r_2} &<& 126.441, & -T_{rw,r_1} -T_{rw,r_2} &<& 94.8101 \\	
	\end{bsmallmatrix},
	\end{align*} whereas for $\mathbf{M}_{da}$
	\begin{align*}
		\begin{bsmallmatrix}
	T_{z_1} & <& 22.3242, & T_{z_2}& <& 22,\\
	T_{rw,r_1}& < &40, & T_{rw,r_2} &< &40,\\
	-T_{z_1} &<& -10.1225,& -T_{z_2}&<& -5.38875,\\
	-T_{rw,r_1} &<& 109.275, & -T_{rw,r_2} &<& -11.1512, \\
	T_{z_1} + T_{z_2}& <& 44, & T_{z_1}+ T_{rw,r_1} &<& 62,\\
	T_{z_1}  + T_{rw,r_2} &<& 62, & T_{z_1} -T_{z_2} &<& 9.50273,\\
	T_{z_1}  - T_{rw,r_1}  &<& 119.446,& T_{z_1} -T_{rw,r_2} &<& 1.20718,\\
	-T_{z_1}  + T_{z_2}  &<& 7, &	-T_{z_1}  + T_{rw,r_1} &<& 25, \\
	-T_{z_1}  + T_{rw,r_2} &<& 25,&-T_{z_1} -T_{z_2} &<& -15.5113, \\
	-T_{z_1} -T_{rw,r_1} &<& 99.1041,&	-T_{z_1} -T_{rw,r_2} &<& -21.2737,\\
	T_{z_2}  + T_{rw,r_1}& <& 62, & T_{z_2}  + T_{rw,r_2} &<& 62,\\
	T_{z_2}  -T_{rw,r_1} &<& 117.336, & T_{z_2}  -T_{rw,r_2} &<& -5.51993,\\
	-T_{z_2}  + T_{rw,r_1} &<& 25, &-T_{z_2}  + T_{rw,r_2} &<& 25,\\
	-T_{z_2}  -T_{rw,r_1} &<& 103.886,&-T_{z_2}  -T_{rw,r_2} &<& -16.5399,\\
	T_{rw,r_1}+ T_{rw,r_2} &<& 80, &	T_{rw,r_1} -T_{rw,r_2} &<& 10,\\
	-T_{rw,r_1}+ T_{rw,r_2} &<& 129.684, & -T_{rw,r_1} -T_{rw,r_2} &<& 98.1239. \\	
	\end{bsmallmatrix}
	\end{align*}
 }

\subsubsection*{Probabilistic reachability analysis using FAUST$^2$}
\label{app:cs:1:faust}
To solve this problem and  achieve low errors within a computationally feasible time frame we abstract the model $\mathbf{M}_s$   
such that 
\begin{align*}
x^T& = [T_{z_1}\;T_{z_2}]^T
\end{align*}
(the states of interest and we fix the rest of the state to steady-state). Based on this model, we construct a stochastic kernel in the form of a Gaussian conditional distribution, $\N(f(x,u),\Sigma)$ where \begin{align*}
f(x,u) = Ax+Bu +Q_d
\end{align*} (cf.~\eqref{mod:s}) and 
\begin{align*}
\Sigma &= diag(\Delta[\sigma_{z_1}^2, \sigma_{z_2}^2])
\end{align*}.
The input space of the process is set to $[15\;22]$, while the safe-set is defined as 
\begin{align*}
\begin{bmatrix}
19.5 & 19.5\\
20.5 & 20.5\\
\end{bmatrix}.
\end{align*}

We select the \texttt{PCTL Safety} option in FAUST$^2$ to compute the analysis. The model is abstracted based on adaptive partitioning, with a partitioning error of $\epsilon = 10^{-5}$,
to obtain the discretised transition kernel. 
The maximal safety probability for each partition is computed, based on the transition kernel, recursively over the whole time horizon $N=6$.

\subsection*{Discrete-time models for two-zone heating setup with large number of continuous variables}
\label{app:cs:2}
We present the corresponding system matrices of the model $\mathbf{M}_{c}$ described in Subsection~\ref{subsec:cs:2} as: 
\begin{align*}
&
A_c=\begin{bsmallmatrix}
0.9998  &	6.54e-9&	2.23e-5	& 2.23e-5&	2.23e-5&	4.88e-14 &	4.88e-14\\
5.739e-9&	0.9998 &	4.27e-14& 4.27e-14&	2.23e-5&	2.23e-5	&	2.23e-5\\
0.0005  &	1.27e-12&	0.9989	& 6.54e-9&	6.54e-9&	7.13e-18&	7.13e-18\\
0.0005  &	1.27e-12&	6.54e-9	& 0.9989&	6.54e-9&	7.13e-18&	7.13e-18\\
0.00051 &	0.00058&	5.73e-9	& 5.73e-9&	0.9989 &	6.54e-9	&	6.54e-9\\
1.11e-12&	0.00058&	6.25e-18& 6.25e-18&	6.54e-9&	0.9989	&	6.54e-9\\
1.11e-12&	0.00058&	6.25e-18& 6.25e-18&	6.54e-9&	6.54e-9	&	0.9980\\
\end{bsmallmatrix},
&
B_c=\begin{bsmallmatrix}
0.000122\\
0.000122\\
3.58e-8\\
3.58e-8\\
6.72e-8\\
3.58e-8\\
3.58e-8\\
\end{bsmallmatrix},\\
&F_c=\begin{bsmallmatrix}
1.027e-8&	5.734e-9&7.31e-9&	2.71e-15&	0.0013& 0.0014\\
1.91e-7 &	5.73e-9&1.39e-17&	1.24e-6 &	0.0021& 0.0022 \\
2.00e-12&	0.0005&	2.13e-12&	3.96e-19&	3.84e-7&3.84e-7\\
0.0009  &	1.11e-12&	2.13e-12&	3.96e-19&	3.84e-7&3.84e-7\\
3.90e-11&	2.09e-12&	1.87e-12&	3.63e-10&	9.78e-7&9.78e-7\\
3.72e-11&	0.00051&	2.042e-21&	3.63e-10&	6.41e-7&6.41e-7\\
0.01708	&1.11e-12&	2.04e-21&	3.63e-10&	6.40e-7&6.41e-7\\
\end{bsmallmatrix},
&
Q_c=\begin{bsmallmatrix}
0.2482\\
-0.0055\\
0.1270\\
0.0201\\
0.0145\\
0.0144\\
0.0145\\
\end{bsmallmatrix}.
\end{align*}
We model the disturbances as random external effects affecting the room temperature dynamics as 
$T_{out}[k] \sim \N(9,1),~T_{hall}[k] \sim \N(15,1),\; 
	CO_{2_i}[k] \sim \N(500,100), i \in \{1,2\}, \;
T_{rw,r_i}[k] \sim \N (35,5), i \in \{1,2\}$.

Next, we present the abstract models $\mathbf{M}_{c_a},a=\{4,\dots,1\}$ taking the same form as \eqref{eqn:Mc} with 
\begin{table}[h!]
	\centering
	\resizebox{0.6\columnwidth}{!}{
	\begin{tabular}{|c|l|l|}
		\hline	
		\textbf{Model} & $\mathbf{x}_c^T$ & $\mathbf{d}_c^T$ \\ \hline \hline   
		$\mathbf{M_{c_4}}$ & $[T_{z_1}\; T_{w_5}\;T_{w_2}\;T_{w_7}]^T$& 	 $[T_{out}\;T_{hall}\;CO_{2_1}\;T_{rw,r_1}\;T_{z_2}]^T$\\ \hline	
		$\mathbf{M_{c_3}}$ & $[T_{z_1}\; T_{w_5}\;T_{w_2}]^T$& $[T_{out}\;T_{hall}\;CO_{2_1}\;T_{rw,r_1}]^T$\\ \hline	
		$\mathbf{M_{c_2}}$ & $[T_{z_1}\; T_{w_2}]^T$& $[T_{out}\;CO_{2_1}\;T_{rw,r_1}]^T$\\ \hline	
		$\mathbf{M_{c_1}}$ & $[T_{z_1}]^T$&$[T_{out}\;CO_{2_1}\;T_{rw,r_1}]^T$\\ \hline	
	\end{tabular}}
\end{table}

and, $T_{z_2}$ is $\sim \N(20,1)$. For $\mathbf{\mathbf{M}_{c_4}}$ we have 
\begin{align*}
&
A_{c_4}=\begin{bsmallmatrix}
0.9998 & 2.23e-5&	2.23e-5&	2.23e-5\\
0.00058& 0.9989 &	6.54e-9&	6.54e-9\\
0.00058&6.54e-9&	0.9989 &	6.54e-9\\
0.00051&5.73e-9&	5.73e-9&	0.9989\\
\end{bsmallmatrix},~
B_{c_4}=\begin{bsmallmatrix}
	0.00012\\
	3.5859e-8\\
	3.5859e-8\\
	3.1424e-8\\
	\end{bsmallmatrix},\\
&
	F_{c_4}=\begin{bsmallmatrix}
	1.02e-8&	5.73e-9&	7.31e-9&	0.0013&	6.54e-9\\
	2.00e-12&	0.0005&	2.13e-12&	3.84e-7&	1.27e-12\\
	0.0009& 1.11e-12&	2.13e-12&	3.84e-7&	1.27e-12\\
	1.75e-12&	9.79e-13&	1.87e-12&	3.37e-7	&0.00058\\
	\end{bsmallmatrix},~
	Q_{c_4}=\begin{bsmallmatrix}
	0.2482\\
	0.1270\\
	0.0145\\
	0.0145\\
	\end{bsmallmatrix},
\end{align*} while for $\mathbf{M}_{c_3}$,
\begin{align*}
&
	A_{c_3}=\begin{bsmallmatrix}
	0.9998&	2.23e-5&	2.23e-5\\
	0.00058&	0.9989&	6.54e-9\\
	0.00058&	6.54e-9&	0.9980\\
	\end{bsmallmatrix},~
	B_{c_3}=\begin{bsmallmatrix}
	0.000122\\
	0.000122\\
	3.58e-8\\
	\end{bsmallmatrix},\\
&
	F_{c_3}=\begin{bsmallmatrix}
	6.29e-9&	5.73e-9&	7.31e-9&	0.0013\\
	1.22e-12&	0.00051&	2.13e-12&	3.84e-7\\
	0.00056&	1.11e-12&	2.13e-12&	3.84e-7\\
	\end{bsmallmatrix},~
Q_{c_3}=\begin{bsmallmatrix}
	0.2482\\
	0.1270\\
	0.0145\\
	\end{bsmallmatrix}.
\end{align*}
For $\mathbf{M}_{c_2}$ the system matrices are 
\begin{align*}
&
A_{c_2}=\begin{bsmallmatrix}
	0.9998&	2.237e-5\\
	0.00058&	0.9989\\
	\end{bsmallmatrix},~
B_{c_2}=\begin{bsmallmatrix}
0.00012\\
3.58e-8\\
	\end{bsmallmatrix},\\
&
	F_{c_2}=\begin{bsmallmatrix}
1.027e-8&	7.31e-9&	0.0013\\
0.00091&	2.13e-12&	3.84e-7\\
	\end{bsmallmatrix},~
	Q_{c_2}=\begin{bsmallmatrix}
	0.2482\\
	0.1270\\
	\end{bsmallmatrix}.
\end{align*}
and  $\mathbf{M}_{c_1}$ is described using 
\begin{align*}
&
A_{c_1}=\begin{bsmallmatrix}
0.9998\\
	\end{bsmallmatrix},
B_{c_1}=\begin{bsmallmatrix}
	0.000122\\
	\end{bsmallmatrix},
	&
	F_{c_1}=\begin{bsmallmatrix}
	6.31e-5&	7.31e-9&	0.0013\\
	\end{bsmallmatrix},~
	Q_{c_1}=\begin{bsmallmatrix}
	0.2482\\
	\end{bsmallmatrix}.
\end{align*}
\newpage
\subsection*{Models for heating system setup with multiple switching controls}
\label{app:cs:3}

To construct the hybrid model we use 
$T_{w_{ss}} = 18^oC,~T_{out}= 25^oC,~T_{SP} = 20^oC,$ 

\noindent$
\delta= 1^oC,~ \delta_2= 2^oC,~\delta_3 = 3^oC, ~ \delta_4=4^oC,~\delta_5 = 5^oC,\;
m_{a,med}= 10m^3/hr,  m_{a,high}=15m^3/hr~CO_{2_{ss}}= 500ppm$.
The continuous dynamics are composed via the following simplifying assumptions:  
(i) the outside air temperature is fixed $T_{out} = 25^oC$ and is used to warm up the zone such the temperature set-point $T_{SP}=20^oC$ is maintained;  
(ii)  we fix $CO_{2_{1_{ss}}}$; 
(iii) there is a constant wall temperature $T_{w_{ss}}$; 
(iv) there are no radiators within the zone; and 
(v) there is no process noise. 
The resulting continuous state models for each discrete mode using the corresponding models presented in Table~\ref{tab:cs2:cont}.\\
\begin{table}[h!]
	\centering
	\resizebox{0.8\columnwidth}{!}{
			\begin{tabular}{|c|l|}
				\hline	
				\textbf{Discrete state} & \textbf{Continuous space dynamics} \\ \hline \hline
				$q= (O,-)$ & $\dot{T}_{z_1} = - 0.0116T_{z_1}(t) + 0.2565$ \\
				& $\dot{T}_{sa} =  0.0183T_{z_1}(t) - 0.0183T_{sa}(t)$\\ \hline
				$q= (M,Op)$ & $\dot{T}_{z_1} = - 0.0292T_{z_1}(t) + 0.0176T_{sa}(t) + 0.2565$ \\
				& $\dot{T}_{sa} =  0.0183T_{z_1}(t) -0.0185T_{sa}(t) + 0.005$\\ \hline
				$q= (M,Cl)$ & $\dot{T}_{z_1} =  - 0.0292T_{z_1}(t) + 0.0176T_{sa}(t) + 0.2565$ \\
				& $\dot{T}_{sa} =  0.0183T_{z_1}(t) -0.0183T_{sa}(t)$\\ \hline
				$q= (H,Op)$ & $\dot{T}_{z_1} =  -0.038T_{z_1}(t) + 0.0264Tsa(t) + 0.2565$ \\
				& $\dot{T}_{sa} =  0.0183T_{z_1}(t) -0.0186T_{sa}(t) + 0.0076$\\ \hline			
				$q= (H,Cl)$ & $\dot{T}_{z_1} =  -0.038T_{z_1}(t) + 0.0264Tsa(t) + 0.2565$ \\
				& $\dot{T}_{sa} =  0.0186T_{z_1}(t) -0.0186T_{sa}(t)$.\\ \hline				
			\end{tabular}}
	\caption[Continuous state models for each discrete mode.]{\begin{varwidth}[t]{\linewidth}Continuous state models for each discrete\\ mode.\end{varwidth}}
	\label{tab:cs2:cont}
\end{table}

\subsubsection*{Reachability analysis using SpaceEx}
\label{app:cs:3:reach}

We first implement the hybrid system delineated using Figure~\ref{fig:mods:HS:a} within SpaceEx using SX (the SpaceEx modelling language~\cite{cotton2010spaceex}). 
We further bound the zone temperature to lie between $T_{z_1} \in [10\; 30]$, while supply temperature to lie between $T_{sa} \in [15\; 30]$: corresponding to physically feasible states when BAS is operating in a good condition.  This model file is loaded into SpaceEx and the input model is converted into a flat hybrid automaton representation over which reachability analysis can be performed.
Next, we set-up the configuration file defining (i) the initial states in the form of \texttt{"Tz1==15\&Tsa==20\&loc()==off"}, where \texttt{"off"} refers to the initial state label, (ii) any forbidden states (in our case study there is none), (iii) the time-horizon, and (iv) the direction of the reach sets (we select \texttt{oct}, but similar results are achieved when the direction of reach set was set to \texttt{box}).
Once the configuration file is loaded, the reachability analysis algorithm is run over the defined time horizon, which in our case corresponds to two hours and the reachable sets are generated (cf. Figure~\ref{fig:cs3:reach}).

%

\end{document}